\documentclass[twocolumn,tighten]{aastex631}

\usepackage{xcolor}
\usepackage{CJK}

\newcommand\msun{$\rm M_{\odot}$}

\received{September 1, XXXX}
\revised{October 1, XXXX}
\accepted{November 1, XXXX}

\submitjournal{XXX}

\shorttitle{SN~2023ixf Precursor}
\shortauthors{Dong et al.}
\graphicspath{{./}{figures/}}

\begin{document}

\title{A comprehensive optical search for pre-explosion outbursts from the quiescent progenitor of SN~2023ixf}

\correspondingauthor{Yize Dong}
\email{yizdong@ucdavis.edu}

\author[0000-0002-7937-6371]{Yize Dong \begin{CJK*}{UTF8}{gbsn}(董一泽)\end{CJK*}}
\affiliation{Department of Physics and Astronomy, University of California, 1 Shields Avenue, Davis, CA 95616-5270, USA}


\author[0000-0003-4102-380X]{David J.\ Sand}
\affiliation{Steward Observatory, University of Arizona, 933 North Cherry Avenue, Tucson, AZ 85721-0065, USA}

\author[0000-0001-8818-0795]{Stefano Valenti}
\affiliation{Department of Physics and Astronomy, University of California, 1 Shields Avenue, Davis, CA 95616-5270, USA}

\author[0000-0002-4924-444X]{K.\ Azalee Bostroem}
\altaffiliation{LSSTC Catalyst Fellow}
\affiliation{Steward Observatory, University of Arizona, 933 North Cherry Avenue, Tucson, AZ 85721-0065, USA}

\author[0000-0003-0123-0062]{Jennifer E.\ Andrews}
\affiliation{Gemini Observatory, 670 North A`ohoku Place, Hilo, HI 96720-2700, USA}

\author[0000-0002-0832-2974]{Griffin Hosseinzadeh}
\affiliation{Steward Observatory, University of Arizona, 933 North Cherry Avenue, Tucson, AZ 85721-0065, USA}

\author[0000-0003-2744-4755]{Emily Hoang}
\affil{Department of Physics and Astronomy, University of California, 1 Shields Avenue, Davis, CA 95616-5270, USA}

\author[0000-0003-0549-3281]{Daryl Janzen}
\affiliation{Department of Physics \& Engineering Physics, University of Saskatchewan, 116 Science Place, Saskatoon, SK S7N 5E2, Canada}

\author[0000-0001-5754-4007]{Jacob E.\ Jencson}
\affil{Department of Physics and Astronomy, Johns Hopkins University, 3400 North Charles Street, Baltimore, MD 21218, USA}
\affil{Space Telescope Science Institute, 3700 San Martin Drive, Baltimore, MD 21218, USA}


\author[0000-0001-9589-3793]{Michael Lundquist}
\affiliation{W.~M.~Keck Observatory, 65-1120 M\=amalahoa Highway, Kamuela, HI 96743-8431, USA}

\author[0000-0002-7015-3446]{Nicolas E.\ Meza Retamal}
\affiliation{Department of Physics and Astronomy, University of California, 1 Shields Avenue, Davis, CA 95616-5270, USA}

\author[0000-0002-0744-0047]{Jeniveve Pearson}
\affiliation{Steward Observatory, University of Arizona, 933 North Cherry Avenue, Tucson, AZ 85721-0065, USA}

\author[0000-0002-4022-1874]{Manisha Shrestha}
\affil{Steward Observatory, University of Arizona, 933 North Cherry Avenue, Tucson, AZ 85721-0065, USA}


\author[0000-0002-6703-805X]{Joshua Haislip}
\affiliation{Department of Physics and Astronomy, University of North Carolina, 120 East Cameron Avenue, Chapel Hill, NC 27599, USA}
\author[0000-0003-3642-5484]{Vladimir Kouprianov}
\affiliation{Department of Physics and Astronomy, University of North Carolina, 120 East Cameron Avenue, Chapel Hill, NC 27599, USA}
\author[0000-0002-5060-3673]{Daniel E.\ Reichart}
\affiliation{Department of Physics and Astronomy, University of North Carolina, 120 East Cameron Avenue, Chapel Hill, NC 27599, USA}



\begin{abstract}

We perform a comprehensive search for optical precursor emission at the position of SN~2023ixf using data from the DLT40, ZTF and ATLAS surveys.
By comparing the current data set with precursor outburst hydrodynamical model light curves, we find that the probability of a significant outburst within five years of explosion is low, and the circumstellar material (CSM) ejected during any possible precursor outburst is likely smaller than $\sim$0.015\msun. By comparing to a set of toy models, we find that, if there was a precursor outburst, the duration must have been shorter than $\sim$100 days for a typical brightness of $M_{r}\simeq-9$ mag or shorter than 200 days for $M_{r}\simeq-8$ mag; brighter, longer outbursts would have been discovered. 
Precursor activity like that observed in the normal type II SN~2020tlf ($M_{r}\simeq-11.5$) can be excluded in SN~2023ixf. 
If the dense CSM inferred by early flash spectroscopy and other studies is related to one or more precursor outbursts, then our observations indicate that any such outburst would have to be faint and only last for days to months, or it occurred more than five years prior to the explosion.
Alternatively, any dense, confined CSM may not be due to eruptive mass loss from a single red supergiant (RSG) progenitor.  Taken together, the results of SN~2023ixf and SN~2020tlf indicate that there may be more than one physical mechanism behind the dense CSM inferred around some normal type II SNe.

\end{abstract}



\section{Introduction} \label{sec:intro}
Red supergiant (RSG) stars with zero-age main sequence masses in the range $\sim$$8-17$ \msun{} can explode as Type II supernovae (SNe) \citep{VanDyk2003,Smartt2009,Smartt2015,VanDyk2017,VanDyk2023}. Early SN observations provide hints about the circumstellar environment around the progenitor star just prior to explosion.  For instance, spectroscopic observations within days of explosion show narrow `flash' recombination lines in a significant fraction of normal SNe II, which quickly disappear after several days \citep[e.g..][]{Khazov2016,Bruch2021,Bruch2022}.  A standard interpretation is that these lines signal dense, confined CSM that has been ionized by the shock breakout \citep[e.g.][]{Yaron2017} or ejecta interaction \citep[e.g.][]{Leonard2000,Smith2015,Terreran2022}. Meanwhile, the fast rise of type II SN light curves has also been interpreted as a sign of dense CSM around the progenitor star, as indicated by hydrodynamic modeling \citep{Morozova2017,Morozova2018}.  Between 40--70\% of standard type IIP SNe show evidence of dense CSM around their progenitor stars \citep[e.g.][]{Forster2018,Morozova2018,Bruch2022}.

The dense CSM around the progenitor requires intense mass loss, equivalent to $\sim$10$^{-4}$--10$^{-2}$ M$_{\odot}$ yr$^{-1}$, in the months to years leading up to explosion, much higher than the mass loss due to the normal stellar winds of RSGs. However, how and when this enhanced mass loss occurs is still a mystery. Some of the possible mass-loss mechanisms are mass ejection driven by wave transport \citep{Quataert2012, Shiode2014, Fuller2017, Morozova2020}, common envelope interaction with a compact object \citep{Chevalier2012, Soker2013}, and dynamical instability associated with turbulent convection in the core \citep{Smith2014}.

One direct method to constrain very late-stage mass-loss mechanisms of SN progenitors is searching for signs of pre-explosion activity or precursor emission. 
Precursor emission has been observed in many SNe IIn \citep[e.g.,][]{Mauerhan2013, Ofek2013Natur.494...65O, Tartaglia2016, Pastorello2013,Pastorello2018} and statistical studies on a sample of SNe IIn also support the idea that most experienced outbursts prior to exploding \citep{Ofek2014, Strotjohann2021}. In contrast, pre-explosion activity in normal Type IIP/L has only been seen in SN~2020tlf, where excess emission is observed $\sim$130 days prior to and all the way up until the ultimate SN explosion \citep{Jacobson-Galan2022}.
Based on the pre-explosion images of four Type IIP/L SN progenitors, \cite{Johnson2018MNRAS.480.1696J} found that the probability that their progenitors had extended outbursts after oxygen ignition is low. However, they could not exclude short outbursts on the time-scale of months from their data. 



Recently, there has been theoretical research on the morphology of precursor light curves of Type IIP/L SNe.
\cite{Davies2022} constructed model spectra of the precursor emission for different mass-loss scenarios. They suggested that the precursor outburst likely occurs within one year of the explosion and would be optically bright for a few days with $M_{R}\simeq-8.5$, accompanied by intense mass loss.
In addition, RSGs can be very faint in the optical right before explosion due to the cooling of their surfaces and an increase of the molecular opacity \citep{Davies2022}.
\cite{Tsuna2023} modeled precursor outbursts by injecting energy into the base of the RSGs' hydrogen envelopes, and explored the corresponding observational light curves. They found that these outbursts can last for hundreds of days, with a peak brightness of $\sim$$-$8.5 to $-$10 mag in the $R$ band, depending on the amount of energy injected. These kinds of precursors are usually too faint to be detected by most ongoing wide-field surveys. However, if a Type IIP/L SN explodes in a very nearby galaxy, its precursor activity can be used as an early warning of the explosion.

In this paper, we present optical pre-explosion monitoring data at the position of SN~2023ixf, a type II SN that exploded in the very nearby galaxy M101 (also known as the Pinwheel Galaxy). The SN displayed strong flash features indicative of dense, confined CSM around the progenitor star \citep{Yamanaka2023,JacobsonGalan2023,Smith2023,bostroem2023early,Teja2023}. Given the proximity of SN~2023ixf and the wealth of available pre-explosion data, it provides an excellent opportunity to link the signatures of CSM in the SN data to one or more pre-explosion events.  Pre-explosion photometry was gathered from several time domain programs: the Distance Less Than 40 Mpc \citep[DLT40,][]{Tartaglia2018} survey, the Zwicky Transient Facility \citep[ZTF,][]{Bellm2019,Graham2019}, and the Asteroid Terrestrial-Impact Last Alert System (ATLAS, \citealt{Tonry2011,Tonry2018,Smith2020}). The pre-explosion observations span about 3, 5 and 6 years prior to the explosion of SN~2023ixf for DLT40, ZTF and ATLAS, respectively. The high-cadence observations enable us to put strong constraints on any precursor outbursts or other  activities. 

The pre-explosion observations at the position of SN~2023ixf, and associated photometric limits, are described in Section~\ref{sec:observations}. We use these photometric limits from multiple surveys to constrain the duration and brightness of any pre-explosion outbursts in Section~\ref{sec:discussion}, using both toy-model outbursts and those derived from hydrodynamic models. We also discuss our outburst constraints in the context of other evidence for dense, confined CSM in SN~2023ixf and other normal core collapse SNe.
Finally, we present our conclusions in Section~\ref{sec:conclusion}.

\begin{figure*}
\includegraphics[width=1.\linewidth]{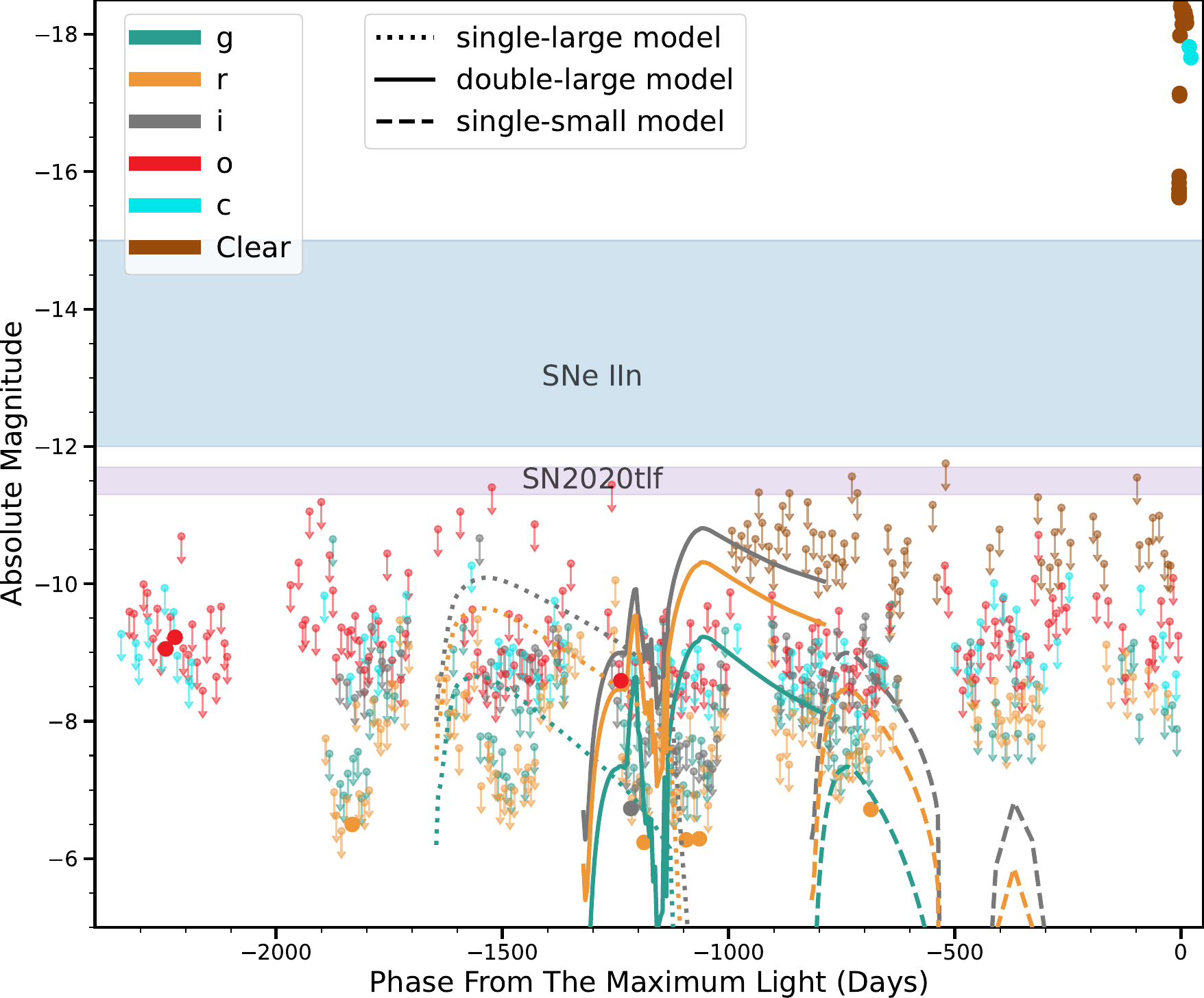}
\caption{Limits on the pre-explosion activity of SN~2023ixf. The Clear filter is calibrated to the $r$ band. A few precursor models from \cite{Tsuna2023} are plotted here for reference. The phases of the models are arbitrarily chosen. The typical brightness of precursor emission in SNe IIn is shown in the blue area. The brightness of precursor emission seen prior to SN~2020tlf is indicated by the purple area. The limits of our data set are generally deeper than the precursor ourbursts observed in Type IIn SNe and in the normal Type IIP SN~2020tlf.
\label{fig:precursor_limit}}
\end{figure*}

\section{Data set}
\label{sec:observations}
SN~2023ixf was discovered on 2023 May 19 in the Pinwheel Galaxy \citep{itagaki_transient_2023} and was classified as a Type II SN \citep{perley_transient_2023}. The distance to SN~2023ixf is only 6.85~Mpc ($\mu$ = 29.18~mag) \citep{Riess2022}, providing an unique opportunity to study a Type II SN in great detail. Following \cite{bostroem2023early}, we adopted a Milky Way extinction of $E(B-V)=0.0077$ mag \citep{Schlafly2011} and a host extinction of $E(B-V) = 0.031$ mag \citep{Smith2023}, as well as $R_V=3.1$.
In this section, we present the pre-explosion data of SN~2023ixf taken by DLT40, ZTF and ATLAS. We also examined the pre-explosion data taken by the All-Sky Automated Survey for Supernovae (ASAS-SN, \citealt{Shappee2014, Kochanek2017PASP..129j4502K}). However, since the survey 
is $\sim$2 mag shallower than the other surveys considered, we did not include the ASAS-SN data in our analysis.
\subsection{DLT40 observations}
The DLT40 survey is a sub-day-cadence SN search \citep{Tartaglia2018,Yang2019}, targeting prominent galaxies within 40 Mpc with the aim of finding about 10 very young and nearby SNe per year. DLT40 has been monitoring M101 since 2020 using the PROMPT-USask 0.4m telescope at Sleaford Observatory, Canada.  These observations resulted in 264 frames taken in the Clear band, with an average time between two adjacent images of $\sim$3.7 days.
Each image has an exposure time of 45s and a field of view of $10'\times10'$.

Before doing any analysis, all the available images were visually inspected, and those of bad quality were removed from the sample.  
A deep template was made using SWarp \citep{Bertin2002} with images taken between 2020-05-12 and 2020-08-31. The rest of the images are stacked in windows of 10 days, and image subtraction against the template was done using HOTPANTS \citep{Becker2015}. 
Aperture photometry was done on difference images to search for any precursor emission at the position of SN~2023ixf.  
For aperture photometry, we adopt an aperture 2 times the FWHM of the image, a signal-to-noise threshold of 3 for source detections and a signal-to-noise threshold of 5 for computing upper limits, following \cite{Masci2011ComputingFU}.
The final Clear-band aperture photometry was performed in a Python-based pipeline and was calibrated to the $r$-band using the APASS catalog. 
This process resulted in a median limiting magnitude of $r$$\sim$$-$10.6 mag.

\subsection{ZTF observations} \label{sec:ztf_obs}
ZTF is a time-domain survey using the Palomar 48-inch Oschin telescope at Palomar Observatory \citep{Bellm2019,Graham2019}. ZTF observes the whole visible sky from Palomar in the $g$ and $r$ filters every two to three nights, and although there is both a public and private portion of the survey, both components are released at regular intervals.
The position of SN~2023ixf had been observed by ZTF for over 5 years before the SN explosion. There are 1092, 1152, and 345 frames taken in the $g$, $r$, and $i$ filters, respectively. The average time between two adjacent images is $\sim$1.7 days for the $g$ band, $\sim$1.6 days for the $r$ band, and $\sim$3.6 days for the $i$ band.
We obtained forced photometry from the template-subtracted images using the ZTF Forced Photometry Service \cite[ZFPS;][]{Masci2023}. Following \cite{Masci2011ComputingFU}, we adopted a signal-to-noise threshold of 3 for the source detection and a signal-to-noise of 5 for computing the upper limit. Bad-quality data were removed following the description in \cite{Masci2023}. We also removed  epochs that have status code 56 to avoid the impact of bad or blank pixels. The single-epoch flux measurements were combined in 10-day time bins following the method described by \cite{Masci2023}. The median limiting magnitudes are $\sim$$-$7.9 mag in $g$ band, $\sim$$-$7.8 mag in $r$ band, $\sim$$-$8.5 mag in $i$ band.

\subsection{ATLAS observations}
ATLAS is an all-sky daily-cadence survey, using two filters, orange ($o$) and cyan ($c$), similar to Pan-STARRS filters $r+i$ and $g+r$, respectively. For over six years prior to the SN explosion, ATLAS had collected 1787 images in the $o$ band and 475 images in the $c$ band. The average time between two adjacent images is $\sim$1.3 days for the $o$ band and $\sim$5.0 days for the $c$ band.
We obtained forced photometry at the supernova position from the ATLAS forced photometry server \citep{Shingles2021}. The single-epoch flux measurements have been stacked in 10-day bins following \cite{Young2022} to reach a deeper limit. 
The median depths we can reach are $\sim-$9.2 mag in $o$ band and $\sim-$8.9 mag in $c$ band.

\begin{figure}
\includegraphics[width=1.\linewidth]{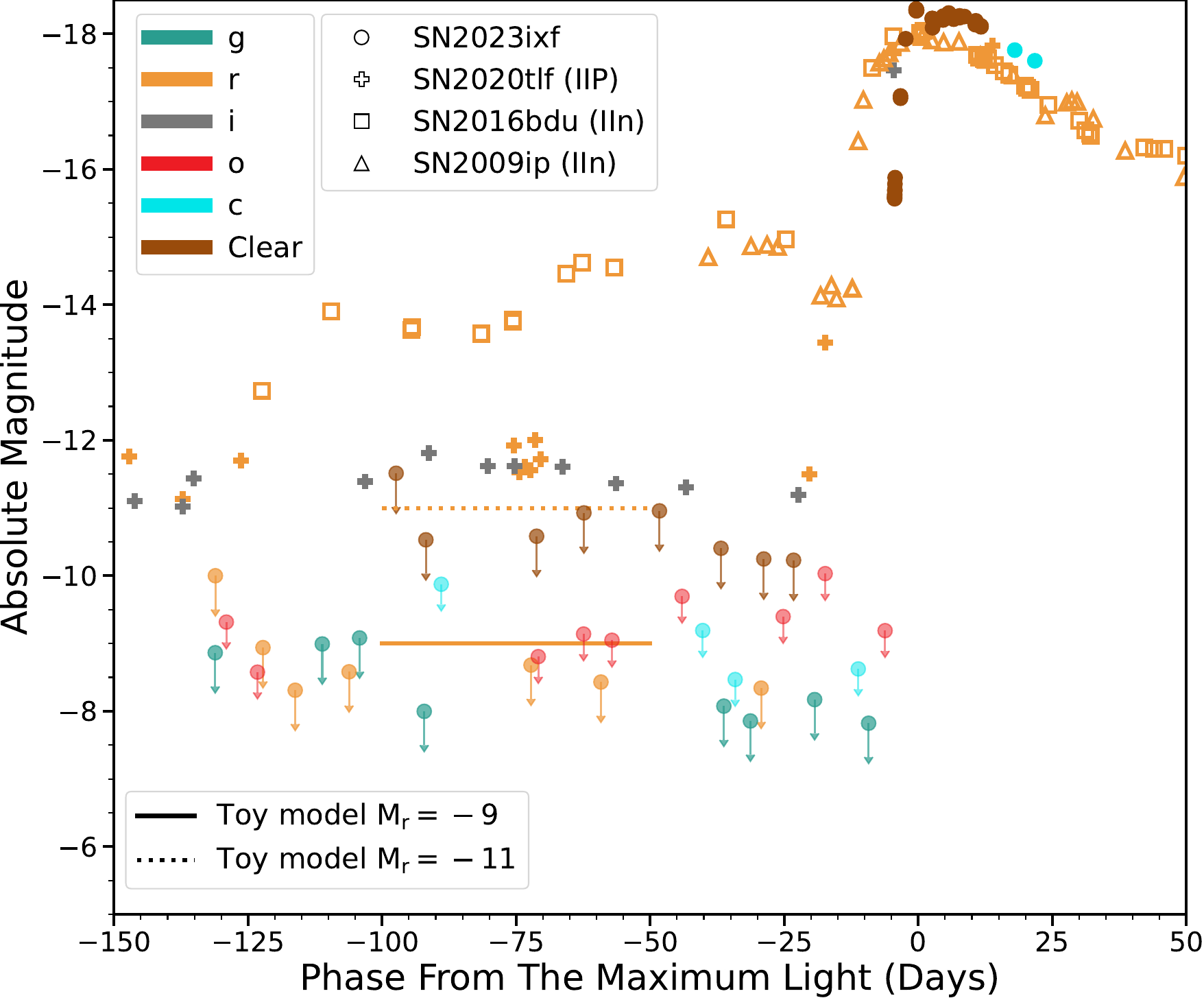}
\caption{A zoomed-in section of Figure \ref{fig:precursor_limit} close to the explosion. The precursor light curves of SN~2020tlf \citep[Type IIP,][]{Jacobson-Galan2022}, SN~2009ip \citep[Type IIn;][]{Mauerhan2013}, and SN~2016bdu \citep[Type IIn;][]{Pastorello2018} are overplotted for comparison. Toy models with a duration of 50 days and luminosities of $M_r = -9$~mag and $-11$~mag are also shown.
\label{fig:precursor_limit_zoomin}}
\end{figure}

\begin{figure}
\includegraphics[width=1.\linewidth]{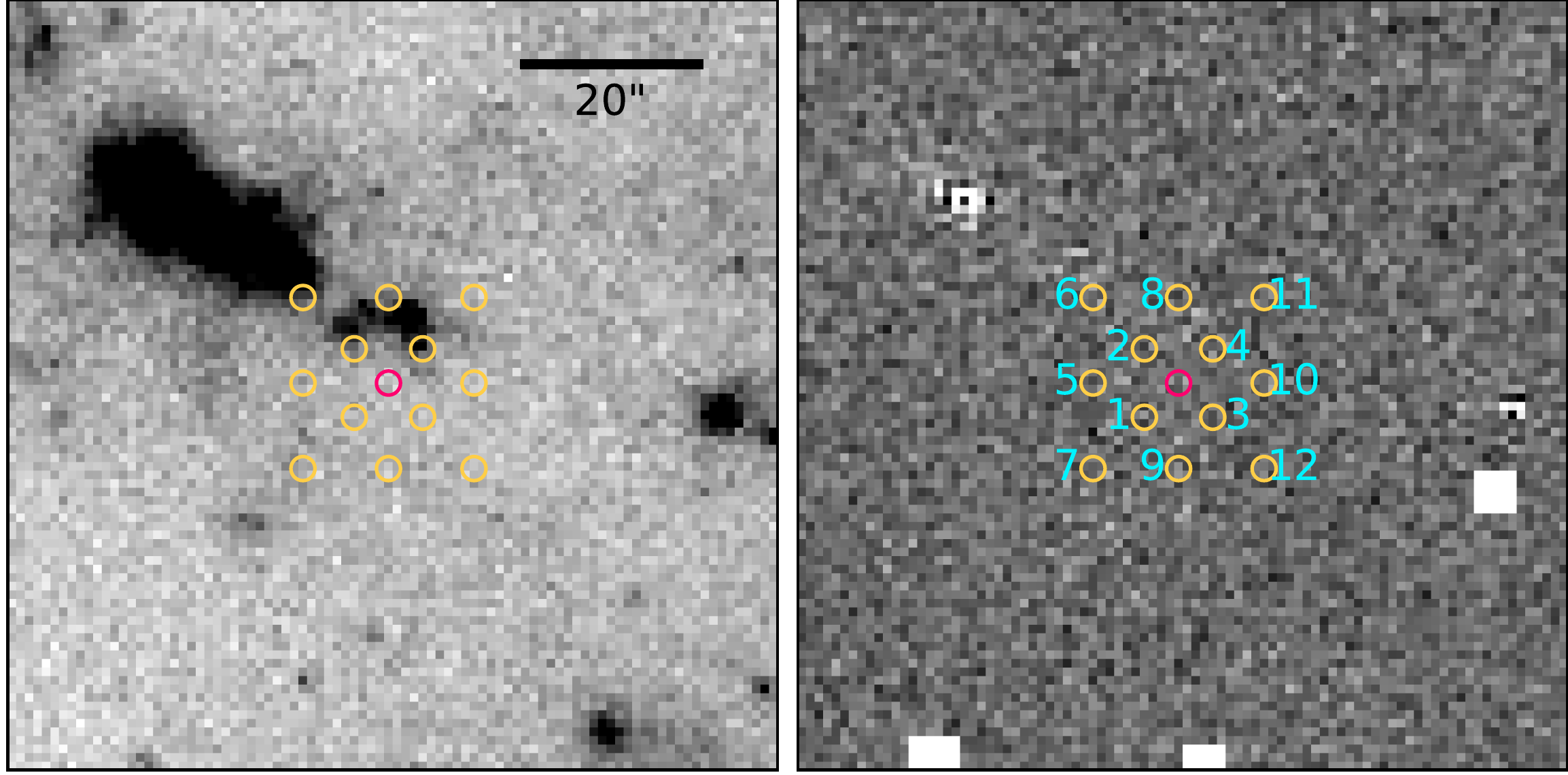}
\caption{ZTF science and difference images at the position of SN~2023ixf before the explosion. The images are downloaded from the ZTF Science Data System \citep{Masci2019} for illustrative purposes. The magenta circle marks the position of the SN. The 12 orange circles show a grid of sample positions around the SN. We performed forced photometry on these sample positions in addition to the SN location to check the reliability of the detections. 
\label{fig:sample_pixel}}
\end{figure}
 
\subsection{Spurious detections} \label{sec:spurious_det}
All the stacked measurements and limits are shown in Figure \ref{fig:precursor_limit}, with a zoom-in around the time of SN~2023ixf's explosion in Figure \ref{fig:precursor_limit_zoomin}.
There are a handful of epochs, both in ZTF and ATLAS, which have reported fluxes larger than 3$\sigma$ and thus are marked as detections (Figure~\ref{fig:precursor_limit}). We have listed the details of these epochs in Table \ref{tab:det}. In all cases, the signal-to-noise ratio of these observations are slightly higher than 3 but smaller than 4. 
In addition, none of these pre-explosion detections are consecutive in time; they are bracketed by non-detections of similar depth.  For this reason, they are likely not true detections of precursor variability.
Given the hundreds of epochs examined, it is expected that some detections at this level would occur, even if they do not indicate true pre-explosion variability. 
Assuming the noise is Gaussian, the number of such data points would be 1 for ZTF and ATLAS, respectively. Additionally, the spurious detections could be potentially due to image reduction issues and unsatisfactory weather conditions.

To further examine the reliability of the detections, we chose 12 positions around the SN position, separated by $\sim$4 to 10 pixels, corresponding to $\sim$4 to 10 arcsecs (illustrated in Figure \ref{fig:sample_pixel}). We then performed forced photometry on these sample positions in an identical manner as we did at the SN position.  By itself, this grid of positions around the SN is unrelated to any transient, but with their close proximity to SN~2023ixf, we can use the sample positions to gauge the rate of random and low signal-to-noise detections in the data. We can also analyze detections that are at similar epochs but different spatial positions (perhaps indicating larger scale image artifacts in data from that time period).

For the ZTF data, we found that there are also several low significance detections in the $r$ and $i$ bands, in positions 3 and 5 in particular, at similar phases to our nominal detections at the SN position. Likewise for the ATLAS data, we found that there are detections in the $o$ and $c$ bands at similar phases for most of the grid sample positions. Given all of the above, we treat the low-significance detections at the SN position as spurious and do not include them in our analysis.




\begin{figure}
\includegraphics[width=1.\linewidth]{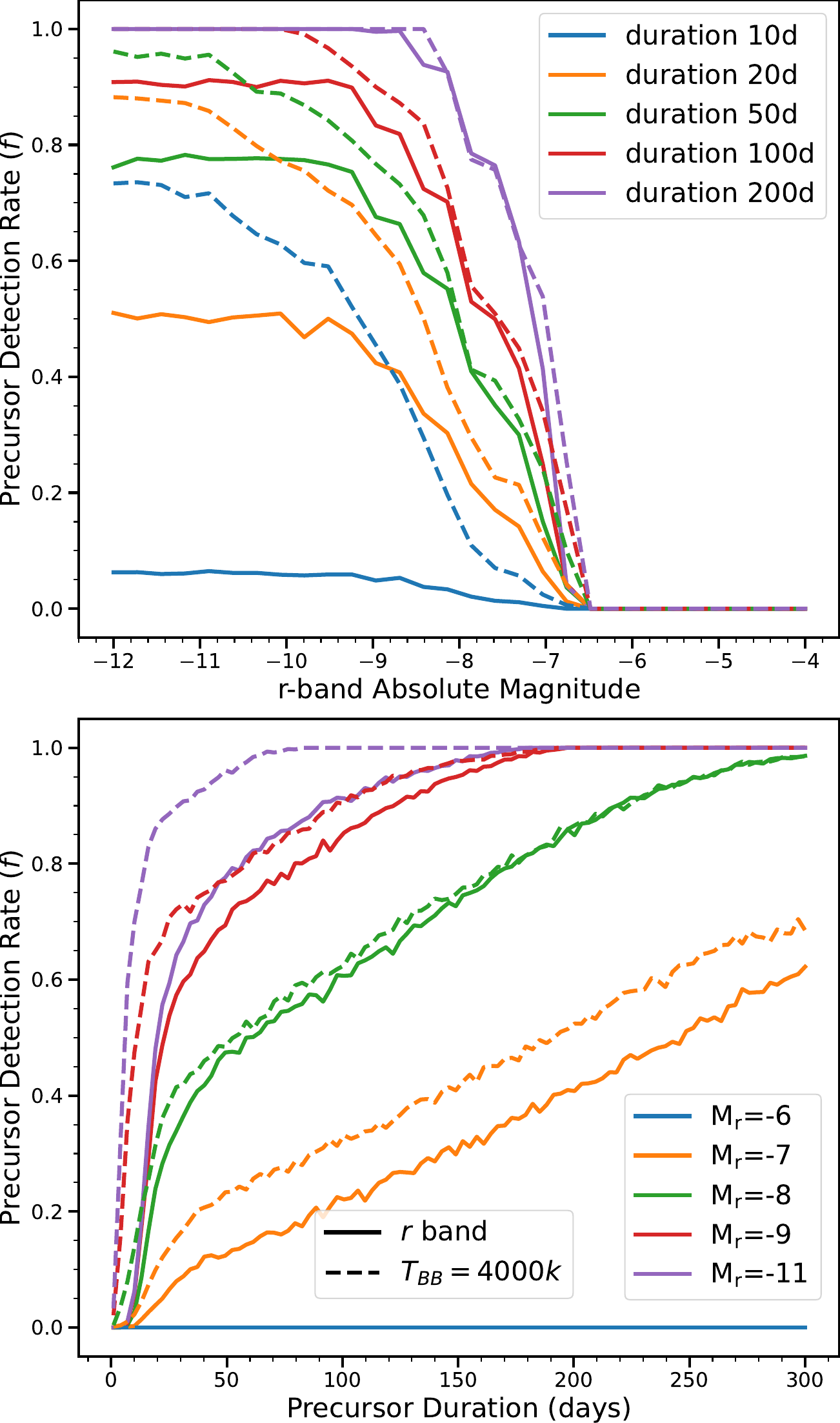}
\caption{Upper: The detection rate of toy-model precursor outbursts given our data set as a function of the r-band absolute magnitude for various outburst durations. The solid line only uses the $r$-band data. The dashed line assumes that the precursor outburst has a blackbody spectrum with a temperature of 4000~K and uses the data from all the filters.
Bottom: The detection rate of a precursor outburst as a function of duration for various outburst absolute magnitudes.
\label{fig:precursor_prob}}
\end{figure}
 
\section{Discussion} \label{sec:discussion}
\subsection{Constraints on the Precursor Activity}
The combined DLT40, ZTF and ATLAS data provide an opportunity to put a strong limit on the brightness and duration of any possible precursor activity in SN~2023ixf. In this section, we discuss the constraint we put on a toy outburst model and the hydrodynamic precursor models of \cite{Tsuna2023}.
\subsubsection{Toy precursor model} \label{sec:toy_model}
We consider a toy burst model with constant brightness and finite duration. Examples of the toy model are shown in Figure \ref{fig:precursor_limit_zoomin}.
For each brightness and duration, we simulated 5,000 outburst light curves during the 5-year period prior to the SN explosion. 
If at one epoch the simulated light curve is brighter than the limit, the data point at that epoch would be marked as a detection. If there are at least two such epochs within 30 days, we will consider the event to be a detected precursor outburst. We note that since we use a signal-to-noise ratio of 5 for calculating the non-detection limit, the `detections' in the experiment presented here would be much more significant than the spurious detections we discussed in section \ref{sec:spurious_det}.
We calculate the detection rate $f$ for each outburst. If $f$ is high, then the probability that we missed a precursor outburst before the explosion of SN~2023ixf is low. Only the $r$-band light curve is used in this analysis. The upper panel of Figure \ref{fig:precursor_prob} shows $f$ as a function of the $r$-band absolute magnitude ($M_{r}$) for various outburst durations (solid lines). Another set of simulations is done by fixing the brightness of the outburst and varying its duration. The result is also shown in solid lines in the bottom panel of Figure \ref{fig:precursor_prob}. 

In order to take advantage of the high-cadence multiband light curves, we assume the precursor has a perfect blackbody spectrum with a constant temperature. We then calculate the magnitude for each filter based on the $M_{r}$. In the outburst precursor model presented by \cite{Davies2022}, the temperature at the progenitor surface can increase by about 1200~K during the outburst. The stellar temperature of the progenitor of SN~2023ixf was estimated to be 3500$^{+800}_{-1400}$~K \citep{jencson2023luminous}. Therefore, we tentatively adopt a blackbody temperature of 4000~K here. 
In the whole data set, if the simulated light curve is brighter than the limit at more than two data points within 30 days, we consider the outburst to be detected by the survey. 
We calculate $f$ using the same method as described above. The results are shown in Figure \ref{fig:precursor_prob} as dashed lines. We note that the blackbody temperature of the precursor in SN~2020tlf is around 5000~K \citep{Jacobson-Galan2022}. As an additional test, we run the same simulation with a temperature of 5000~K and find that this only marginally changes the result.

As presented in Figure \ref{fig:precursor_prob}, for an outburst brighter than $M_{r} = -9$~mag and longer than 100 days, the detection rate of a precursor ($f$) is larger than $\sim$0.9. Such a precursor is very likely to have been detected. In addition, for an outburst brighter than $M_{r} = -8$~mag, $f$ is larger than 0.9 for a precursor that lasts longer than 200 days. Therefore, we conclude that if there was a precursor for SN~2023ixf, the outburst must have been shorter than 100 days if it was brighter than about $M_r=-9$~mag, or shorter than 200 days if it was brighter than about $M_r=-8$~mag. 


\subsubsection{Hydrodynamical precursor models}
\cite{Tsuna2023} modeled precursor outbursts by injecting energy into the hydrogen-rich envelope of a 15 \msun{} RSG progenitor. Two situations were considered in their models, where energy injection occurs once and twice. The models are differentiated by both the number of energy injections and their amounts. To match the passbands of our data set, we produce blackbody spectra based on the temperature and radius from their model results and applied appropriate filter transmission functions. Examples of a few models are shown in Figure \ref{fig:precursor_limit}.

The progenitor mass of SN~2023ixf is estimated to be $11 \pm 2$ \msun{} by \cite{Kilpatrick2023}, 9 to 14 \msun{} by \cite{Neustadt2023}, $17 \pm 4$ \msun{} by \cite{jencson2023luminous}, and $20 \pm 4$ \msun{} by \citet{Soraisam2023}. 
To take the uncertainty of the progenitor mass into account, we added an uncertainty to the brightness of each model.
\cite{Tsuna2023} showed that, for different progenitor masses, the brightness of the precursor light curve 
varies by less than a factor of 2. Therefore, we varied the precursor brightness using a Gaussian distribution centered on the model light curve with a standard deviation of 0.4 mag, which is roughly equivalent to a factor of 2 of luminosity change. 

For each model, we simulated 10,000 light curves by varying the luminosity as described above. These light curves are then randomly distributed in the 5 years of the pre-explosion observations. The energy injection time is at least 200 days prior to the SN explosion, so that all the models can reach the (first) light curve peak before the time of explosion. 
We calculated $f$ using the same method as described in section \ref{sec:toy_model} and listed the results in Table \ref{tab:tsuna_model}.
For all the models, the probability that there was a precursor that was not detected is low. The model with the lowest detection rate ($f$) is the single-small model. This is because this model has the smallest total injected energy, and thus the light curve is faster and dimmer than other models. 
The CSM mass that is ejected in the single-small model is 0.015 \msun, which is the lowest ejected mass among all the models. Given that this model has the lowest detection probability, we can use it to put an upper limit for the mass ejected during the possible pre-explosion outburst.
We conclude that the probability that we would not detect the single-small model precursor is 23 \% within $\sim$5 years prior to the explosion of SN~2023ixf, while this probability is below about 10 \% for all other models.
Therefore, the upper limit of mass ejection during the precursor outburst (if there was one) is around 0.015 \msun.

\begin{deluxetable}{ccc}
\tablenum{1}
\tablecaption{Detection rate of several outburst models\label{tab:tsuna_model}}.
\tablehead{
\colhead{Model} &
\colhead{$f$} &
\colhead{Unbound CSM (\msun)}
}
\startdata
Single-large & 1.0 & 1.2\\
Single-fid & 0.97  & 0.35\\
Single-small & 0.77 & 0.015\\ 
Double-large & 1.0 & 3.6\\
Double-fid & 0.96 & 1.3\\
Double-long & 0.91& 1.2\\
\enddata
\tablecomments{The detection rate of a precursor in our data set for different models from \cite{Tsuna2023}. These models are differentiated by both the number of energy injections and their amounts.}
\end{deluxetable}

\subsection{Comparison with other precursor studies}
Precursor emission has been observed in many Type IIn SNe. These objects likely have extended, dense CSM around their progenitors, which is driving their long-duration narrow-line emission, and which may have been produced by pre-explosion activity in the progenitor \citep{Mauerhan2013, Pastorello2013, Ofek2014, Strotjohann2021}. This pre-explosion emission could be powered by the interaction with the surrounding CSM or the continuum-driven wind, while the underlying triggering mechanism is still uncertain. 
The precursor outbursts in SNe IIn usually have an absolute magnitude between $-15$~mag and $-12$~mag, which is much brighter than the limits in our observations (see Figures \ref{fig:precursor_limit} and \ref{fig:precursor_limit_zoomin}). 

From SN observations, a significant fraction of RSGs are believed to have dense and confined CSM prior to the explosion, which may be because they have experienced intense mass loss before they explode as Type II SNe \citep[e.g., ][]{Morozova2018, Forster2018, Morozova2020, Bruch2022}. However, after analyzing the pre-explosion progenitors of four Type IIP/L SNe using data from the Large Binocular Telescope, \cite{Johnson2018MNRAS.480.1696J} found that these progenitors were quiescent and the probability that they had extended outbursts after oxygen ignition (around 5.4$-$2.6 years before the SN explosion) is low.
To date, precursor emission in a normal Type IIP SN has only been observed in SN~2020tlf \citep{Jacobson-Galan2022}. Both spectroscopic and photometric observations suggest that the progenitor of SN~2020tlf had experienced enhanced mass loss prior to the explosion, and its precursor emission is likely due to the ejection of the outer layer of its progenitor star during final-stage nuclear burning \citep{Jacobson-Galan2022}. The precursor emission in SN~2020tlf is around $-11.5$~mag in the $r$, $i$, and $z$ bands over about 100 days before explosion, which is about 1 mag brighter than our current limit in the DLT40 Clear band and about 3 mag brighter than our limit in the ZTF $g$ and $r$ bands (see Figures \ref{fig:precursor_limit} and \ref{fig:precursor_limit_zoomin}). Therefore, the type of precursor observed in SN~2020tlf can be excluded in SN~2023ixf. 

Multiple flash-spectroscopy studies have found evidence of dense CSM around the progenitor of SN~2023ixf, which requires a mass loss rate of $\rm 10^{-3}-10^{-2} M_{\odot}\rm yr^{-1}$ \citep{JacobsonGalan2023, bostroem2023early}, comparable to or slightly lower than the mass loss rate estimated for SN~2020tlf ($\rm 10^{-2}$ \msun$\rm yr^{-1}$) \citep{Jacobson-Galan2022}.
The lack of similar precursor activity in SN~2023ixf may suggest that there are various physical mechanisms for the formation of dense CSM around the progenitors of normal Type II SNe. 
For instance, \cite{matsuoka2023binary} proposed that the binary interaction in the final evolutionary stage of RSG stars could contribute to the dense CSM around the SN progenitor.
Recently, \cite{Smith2023} found that the CSM around the progenitor of SN~2023ixf is likely asymmetric, which could be a consequence of binary interaction triggered by pre-SN inflation of the RSG during Ne or O burning. In such a binary scenario, eruptive mass loss from a single RSG may not be the driving force behind the dense CSM that we observe.


\subsection{Quiescent progenitor of SN~2023ixf versus enhanced pre-SN mass loss}

After the discovery of SN~2023ixf, many independent studies have suggested that there is dense and confined CSM around the progenitor, implying an enhanced mass loss prior to the explosion.
Recently, by analyzing the early flash spectroscopy of SN~2023ixf, \cite{bostroem2023early} and \cite{JacobsonGalan2023} suggest that, to produce the dense CSM around the progenitor, the mass-loss rate of the progenitor of SN~2023ixf should be around $10^{-3}-10^{-2}$ $\rm M_{\odot} yr^{-1}$. Based on the hard X-ray observations, \cite{Grefenstette2023arXiv230604827G} also found evidence of dense pre-existing CSM, which requires a mass loss rate of $\rm 3\times10^{-4} \rm M_{\odot} yr^{-1}$ before the explosion. In addition, \cite{jencson2023luminous} analyzed the near- and mid-infrared pre-SN imaging of SN~2023ixf and found a lower mass-loss rate of $\rm 10^{-5}-10^{-3}$ $\rm M_{\odot} yr^{-1}$, but it is still higher than the mass-loss rate of typical RSGs in the same luminosity range. They also found that there was no evidence of infrared precursor outbursts up to $\sim$10 days before the explosion.
Furthermore, \cite{Hosseinzadeh2023_2023ixf} found that the very early light curve evolution of SN~2023ixf is inconsistent with shock cooling models, which could be explained by the interaction of dense pre-existing CSM with the SN ejecta, and thus implies an enhanced mass loss before the SN explosion.
However, they also suggested that the unusual light curve behaviour could be due to a pre-explosion eruption around one day before the explosion or even extended duration emission from the shock breakout.

\cite{Neustadt2023} examined imaging from the Large Binocular Telescope ranging from 5600 days to 400 days before the explosion of SN~2023ixf, and they found no progenitor variability in the $R$ band at the level of $10^{3}L_{\odot}$ up to 400 days before the explosion. Due to the sparse coverage of their data, they could not directly exclude short-lived outbursts. However, they argue that short outbursts would still have had a long-lived effect on the dust optical depth, leading to an increase of progenitor luminosity for decades, which they would have observed.
Our data set has a higher cadence up to about 10 days before the SN explosion, but we still found no signs of strong pre-SN activity from the progenitor.
The enhanced pre-SN mass-loss rate of SN~2023ixf derived from the flash spectroscopy and other studies seems in tension with the lack of any precursor emission in SN~2023ixf.

It is possible that the progenitor star had a relatively faint outburst on a time-scale of days to months. The probability that we would have detected this kind of outburst is low. \cite{Davies2022} suggested that the precursor in SNe II-P is likely in the form of abrupt outbursts, in which the progenitor would only be optically bright for a few days before becoming fainter and redder than normal RSGs and ejecting a significant amount of mass into the surrounding space. 
The detection of such an outburst would require higher-cadence observations prior to the time of explosion.
In addition, the flash spectroscopy may not necessarily imply enhanced mass loss. \cite{Kochanek2019MNRAS.483.3762K} suggested that, in a binary system, the shocked boundary layer produced by the collision of winds from two stars generates a high-density CSM around the progenitor, which produces the flash spectroscopy observed in SNe II. 
If the progenitor of SN~2023ixf was in such binary configurations, then enhanced mass loss or an outburst prior to explosion would not be required.  




\section{Conclusions} \label{sec:conclusion}
We used 5 years of pre-explosion data from DLT40, ZTF, and ATLAS to constrain pre-explosion activity in the progenitor of SN~2023ixf. By comparing the data with a toy precursor model, we found that if there was any precursor activity, an outburst with a typical brightness of $M_{r} \simeq -9$ must have had a duration shorter than 100 days, and an outburst with $M_{r} \simeq -8$ must have had a duration shorter than 200 days. 
We also found that the probability that there was a precursor outburst similar to the models of \cite{Tsuna2023} is low, and therefore that the ejected mass prior to the explosion is likely less than 0.015 $\rm M_{\odot}$.
The precursor activity like the outburst observed in SN~2020tlf can be excluded in SN~2023ixf. 
The enhanced mass loss inferred from the early flash spectroscopy and other studies of SN~2023ixf is in some tension with the non-detection of any precursor outbursts. It is possible that there was a faint precursor within five years of the SN explosion that occurred on a time-scale of days to months. Such an outburst would likely not be detected by our current data set. 
Alternatively, the dense, confined CSM may not be due to the enhanced mass loss from a single RSG progenitor. The dense CSM could have, for instance, originated from the interaction of stellar winds of two stars in a binary system.
In summary, it is likely there are various physical mechanisms for the formation of the dense CSM around the progenitors of normal Type II SNe.
In the near future, with the help of the Legacy Survey of Space and Time (LSST) survey, we will be able to put strong constraints on the precursor activities for a sample of Type IIP/L SNe, which will help us better understand the origin of the dense, confined CSM and the very last stages of RSG stellar evolution.

\vspace{10mm}
\section*{Acknowledgements}
We would like to thank Daichi Tsuna for providing the precursor light curve models.

Research by Y.D., and S.V., N.M.R, and E.H. is supported by NSF grant AST-2008108. 
Time domain research by D.J.S.\ is also supported by NSF grants AST-1821987, 1813466, 1908972, \& 2108032, and by the Heising-Simons Foundation under grant \#2020-1864.

This publication was made possible through the support of an LSSTC Catalyst Fellowship to K.A.B., funded through Grant 62192 from the John Templeton Foundation to LSST Corporation. The opinions expressed in this publication are those of the authors and do not necessarily reflect the views of LSSTC or the John Templeton Foundation.

Based on observations obtained with the Samuel Oschin Telescope 48-inch and the 60-inch Telescope at the Palomar Observatory as part of the Zwicky Transient Facility project. ZTF is supported by the National Science Foundation under Grant No. AST-2034437 and a collaboration including Caltech, IPAC, the Weizmann Institute for Science, the Oskar Klein Center at Stockholm University, the University of Maryland, Deutsches Elektronen-Synchrotron and Humboldt University, the TANGO Consortium of Taiwan, the University of Wisconsin at Milwaukee, Trinity College Dublin, Lawrence Livermore National Laboratories, and IN2P3, France. Operations are conducted by COO, IPAC, and UW. The ZTF forced-photometry service was funded under the Heising-Simons Foundation grant
\#12540303 (PI: Graham).

This research made use of Photutils, an Astropy package for detection and photometry of astronomical sources \citep{larry_bradley_2022_6825092}.

%

\facilities{Prompt-USASK, ZTF, ATLAS}


\software{Astropy \citep{astropy13,astropy18}, 
          HOTPANTS \citep{Becker2015},
          Matplotlib \citep{Hunter2007},
          NumPy (https://numpy.org),
          PYRAF,
          Pandas \citep{mckinney-proc-scipy-2010},
          SciPy (https://www.scipy.org),
          SWarp \citep{Bertin2002},
          HOTPANTS \citep{Becker2015},
          Photutils \citep{larry_bradley_2022_6825092},
          the ZTF Forced Photometry Service \citep{Masci2023},
          the ATLAS forced photometry server \citep{Shingles2021}
          }



\appendix
\section{Parameters of Spurious Pre-explosion Detections}
Table \ref{tab:det} presents the epochs that have signal-to-noise ratios larger than 3. All of these detections are below 4$\sigma$ and are not consecutive in time, so they are likely false detections (see discussion in Section \ref{sec:observations}).
\begin{deluxetable*}{cccccccc}
\tablenum{A1}
\tablecaption{Possible pre-explosion detections. \label{tab:det}}
\tablehead{
\colhead{Epoch (days)} &
\colhead{Filter} &
\colhead{JD} &
\colhead{Flux ($\mu$Jy)} &
\colhead{Flux Error ($\mu$Jy)} &
\colhead{S/N} &
\colhead{$N_\mathrm{frame}$} &
\colhead{Source} 
}
\startdata 
$-$2237.7 &$o$& 2457845.5 & 29.88 & 8.66 & 3.45 & 25 & ATLAS \\
$-$2217.2 &$o$& 2457866.0 & 34.88 & 10.48 & 3.33 & 25 & ATLAS \\
$-$1825.7 &$r$& 2458257.5 & 2.80 & 0.84 & 3.33 & 16 & ZTF \\
$-$1232.1 &$o$& 2458851.1 & 19.5 & 6.0 & 3.25 & 4 & ATLAS \\
$-$1210.2 &$i$& 2458873.1 & 3.55 & 1.01 & 3.52 & 11 & ZTF \\
$-$1180.9 &$r$& 2458902.4 & 2.20 & 0.56 & 3.90 & 28 & ZTF \\
$-$1087.7 &$r$& 2458995.6 & 2.28 & 0.65 & 3.49 & 15 & ZTF \\
$-$1058.7 &$r$& 2459024.5 & 2.31 & 0.68 & 3.41 & 20 & ZTF \\
$-$679.8  &$r$& 2459403.5 & 3.42 & 1.11 & 3.08 & 8  & ZTF \\
\enddata
\tablenotetext{\star}{Epoch is measured with respect to the explosion time \citep[JD~2460083.25;][]{Hosseinzadeh2023_2023ixf}. $N_\mathrm{frame}$ refers to the number of individual image measurements that were combined for a given epoch, within the 10-day bin used throughout this work.}
\end{deluxetable*}




\bibliography{precursors}{}
\bibliographystyle{aasjournal}



\end{document}